# Electrostatic repulsion between an uncharged or slightly charged conductor and a point charge

Vladimir P. Savin[1*], Yury A. Koksharov[1,2]

[1]Faculty of Physics, M.V.Lomonosov Moscow State University, Leninskie Gory, Moscow 119991, Russia

[2]Faculty of Material Science, Shenzhen MSU-BIT, Shenzhen, 518179, China

*Corresponding author.

svladimir1999@yandex.ru (V.P.Savin)

**Abstract**

A counterintuitive effect of a repulsion between uncharged and charged conductors is considered. Our numerical calculations prove this effect for various axially symmetric systems of a neutral conductor in an electric field of a point charge. The presence of a cavity in the conductor is very important for the repulsion effect. Our simple analytical model explains this effect in terms of specific spatial positions of induced charges.

## 1. Introduction

Many electrostatics problems of an interaction between conductors were solved [Smythe, 1954; Greenberg, 1948; Jackson, 1999]. Most solutions require specific mathematical techniques suitable only for a few of tasks. If the image method is applicable then analytical solutions exist. The formulas can be very simple, as for a sphere in the field of a point charge [Jackson, 1999], or quite complicate, as for two charged spherical conductors, because of special functions or infinite series [Smythe, 1954; Kolikov, 2012; Lindgren 2016]. The interaction of complex-shaped conductors can be even counterintuitive, as for a conducting uncharged hemisphere in the field of a point charge [Levin 2011, 2011a]. Levin et al. considered a point charge and an electric dipole acting on neutral open hemisphere and neutral conducting plane with a circular hole, respectively, and found an amazing repulsion effect. The repulsion force occurs when the point charge and the dipole are located in some areas near the entrance to the hemisphere or the hole in the plane, respectively. These examples demonstrate the importance of the conductor shape in such problems.

The existence of a repulsive force between a neutral and a charged conductor has led to an interesting discussion of the quantum mechanical Casimir forces [Falling, 2007; Levin, 2010; Grushin, 2011; Wilson, 2015; Bach 2007; Kenneth 2006; Bachas, 2007]. Some authors [Falling, 2007; Levin, 2010; Grushin, 2011; Wilson, 2015] admit the existence of the repulsive Casimir force. The repulsive regime of the Casimir effect is predicted for an elongated metal particle near a metal plate with a hole [Levin, 2010]. The repulsion Casimir effect can also be expected for dielectrics [Crushing, 2011] and Weyl semimetals [Wilson, 2015]. Some authors [Bachas 2007, Kenneth 2006] reject the existence of the repulsive Casimir force regardless of system geometry and the electric properties of materials [Kenneth 2006]. An experiment could resolve this dispute.

However, the Casimir forces are weak and the precision of modern measurements [Klimchitskaya, 2009; Mandey, 2009; Levin, 2010] is still insufficient to detect them. Nevertheless, imagined measurements could be proposed. Wilson suggested at least four ways to demonstrate the repulsive Casimir force including measurements in systems of variously shaped conductors [Wilson, 2015].

Conceivably, the existence of an open cavity in a neutral conductor is favorable for the repulsive force between charged and neutral conductors. To check this assumption, we numerically investigated uncharged conductors of various shapes (hemisphere, truncated sphere, prolate and oblate semiellipsoids of revolution) in the electric field of a point charge. For simplicity sake, we considered only axisymmetric systems.

Our computational approach is based on the representation of axisymmetric thin-walled conductors as a set of a large number of parallel uniformly charged rings. The equilibrium charges of the rings were calculated by using two methods. The first is to minimize electrostatic potential deviations on the conductor surface. The second is to minimize the electrostatic energy satisfying Thomson's theorem [Jackson, 1999].

## 2. Methods

Axially symmetric conductor can be considered as a set of co-axis homogeneous charged rings $C_i$ (i=1, 2,..., N) (Fig.1a). The electric field from the ring $C_i$ at the point $P(\rho,\psi,z)$, where $\rho,\psi,z$ are the polar coordinates (Fig.1b), can be calculated from Eqs.(1), (2) [Zhu 2005]:

$$\vec{E} = \frac{Q_i k_i}{2\pi\sqrt{a_{ri}\rho^3}} \{K(k_i) - \frac{E(k_i)}{1-k_i^2}[1 - \frac{(\rho + a_{ri})k_i^2}{2a_{ri}}]\}\vec{e_\rho} + \frac{Q_i k_i^3 (z-z_i)E(k_i)}{4\pi a_{ri}(1-k_i^2)\sqrt{\rho^3 a_{ri}}}\vec{e_z}; \quad (1)$$

$$\varphi = \frac{2Q_i}{\pi\sqrt{(z-z_i)^2 + (\rho+a_{ri})^2}} K(k_i), \quad (2)$$

where $K(k_i) = \int_0^{\pi/2} \frac{d\beta}{\sqrt{1-k_i^2\sin^2\beta}}$, $E(k_i) = \int_0^{\pi/2}\sqrt{1-k_i^2\sin^2\beta}d\beta$ are the elliptic integrals of the first and second kind, respectively [Press 1992b], $k_i^2 = \frac{4\rho a_{ri}}{(z-z_i)^2+(\rho+a_{ri})^2}$, $Q_i$ is the total charge of $C_i$, $z_i$ is the coordinate of the $C_i$ center.

Equation (2) for points $P(0,0,z)$ on the symmetry axis z ($\rho=0$; $k=0$; $K(0)=\pi/2$) can be simplified:

$$\varphi = \frac{Q_i}{\sqrt{(z-z_i)^2 + a_{ri}^2}} \quad (3)$$

In our methods Eq.(3) was used to calculate the potential $\varphi_q$ produced by a conductor at the point charge location. Equation (1) allows to check vanishing of a tangential component of ***E*** on the conductor's surface. This condition is an additional proof of achieving electrostatic equilibrium

We used two methods of computation of the equilibrium charges. The first minimizes deviations of $\varphi_i$ (i=1, 2,..., N) from the average value $\varphi_0 = (\sum_{i=1}^{N} \varphi_i)/N$, where $\varphi_i$ is the potential at the point $P(a_{ri},0,z_i)$. The second finds the minimum of the total electrostatic energy *W*. Both methods preserve constancy of the conductor total charge *Q*. The minimization procedure for the first method is based on the "lip-frog" algorithm [Snyman, 2018]. This algorithm is well suited for problems where a minimized function contains experimental or numerical noise [Snyman, 2000].

The electrostatic energy $W$ of the system can be written as

$$W = \frac{1}{2}Q\varphi_0 + \frac{1}{2}q\varphi_q, \qquad (4)$$

where $\varphi_q$ is the value of the potential in the point charge location. In the case of the neutral conductor ($Q=0$) Eq.(3) reduces to

$$W = \frac{1}{2}q\varphi_q. \qquad (5)$$

In our approach the potential $\varphi_q$ in Eq.(5) depends only on the charges $Q_i$ of the rings $C_i$ (Fig.1).

The second method is based on Thompson's theorem for the electrostatic energy $W$ of a charged conductor system. Recall Thomson's theorem: the induced charges on conductors always arrange themselves to minimize the total electrostatic energy of the system [Jackson, 1999]. In our model the total electrostatic energy $W$ is a linear function of $Q_i$, i=1, 2,…, N. The conditional extremum of $W$ was found, provided that $Q = \sum_{i=1}^{N} Q_i = const$, by solving a set of linear algebraic equations using the LU decomposition [Press, 1992].

## 3. Repulsion effect for various shaped uncharged perfect conductors in the field of a point charge

### 3a. Sphere and hemisphere

We tested our numerical methods considering problems for which analytical solutions are known. Figure 2 shows the exact and numerical results related to the electrostatic interaction between the point charge and the neutral perfect conductors: (a) closed sphere and (b) open hemisphere. The exact solutions in Fig.2 were obtained by the image method, for the sphere, and the method of three-dimensional inversion transformation, for the hemisphere [Levin, 2011].

The symmetry axis contains the center of a conductor at $z = 0$ and the point charge $q$ in the intervals $z_q \epsilon (R, +\infty)$, for the sphere, or $z_q \epsilon (-R, +\infty)$, for the hemisphere. The dependence $W$ on $z_q$ is monotonous and the interaction is always attractive (no repulsion effect) for the sphere. For the hemisphere $W$ is not monotonous: $W$ decreases as $z_q$ increases in the interval $(0, 0.63R)$ and $W$ increases as $z_q$ increases in intervals $(-R, 0)$ and $(0.63R, +\infty)$ (Fig.2b). The interaction corresponds to the repulsion in the interval $z_q \epsilon (0, 0.63R)$ where $\partial W (zq)/\partial z_q < 0$.

The comparison in Fig.2 of the exact [Levin, 2011] and our numerical results demonstrates the applicability of both calculation methods. One can see that the first method is more accurate. On the other hand, the second method is faster compared to the first. An increase in number of the rings $N$ leads to a decrease of numerical calculation error. For example, the relative error in $W$ at $z_q = 0.63R$ (Fig.2b) decreases from 4% for N=500 down to 1% for N=2000 in the case of the first method. If N>1000, the relative error in $W$ does not exceed 10% even at the worst case, when $z_q<0$.

Figure 2d presents the equilibrium charge distribution on the neutral open hemisphere, if $W$ has the local minimum at $z_q = 0.63R$. The distribution of negative induced charges is noticeably more compact compared to positive charges. The magnitude of the positive charge increases when approaching the pole of the hemisphere. In section 5, this quasi-continuous distribution will be reduced to discrete elements, a circle with a negative charge at z = 0 and a positive point charge at $z = -R$, in order to present a simplified equivalent model of the repulsion effect.

### 3b. Semiellipsoid of revolution

Let $a$, $b$ denote semi-axes of the open semiellipsoid of revolution (Fig.4a). The hemisphere (Fig.2a) is identical to the semiellipsoid if $a=b=R$. It is assumed that $R$ is a unit for all quantities with dimension of length. The semiellipsoid is centered at the origin of coordinates, the semi-axis $a$ is aligned along z-axis. The point charge is located at the point (0, 0, $z_q$). Figure 4b shows the dependencies $W(z_q)$ for different values of $a$ and $b$. Every curve $W(z_q)$ has a local minimum $W_{min}$ at $z_{min}$ and a local maximum $W_{max}$ at $z_{max}$ (see Fig.4b).

If $a/b < 1$ (oblate semiellipsoid), $z_{min}$ increases (Fig.4a) and $|W_{min}|$ decreases (Fig.4b), respectively, as the eccentricity $\varepsilon = \sqrt{1 - a^2/b^2}$ increases. If a/b>1 (prolate semiellipsoid), $z_{min}$ decreases (Fig.4a) and $|W_{min}|$ increases (Fig.4b), respectively, with increasing $\varepsilon = \sqrt{1 - b^2/a^2}$. We can assume that the repulsion effect steadily disappears in the case of compression of an open semiellipsoid. The results in Fig.4 are consistent with the limiting case $a/b \to 0$, when an oblate semiellipsoids gradually transforms to a plane and the local minimum of $W(z_q)$ disappears. This disappearance is realized by a gradual merging of the extremes of $W(z_q)$, as can be seen from Figs.4c and 4d, where the dependencies $z_{min} - z_{max}$ and $W_{max} - W_{min}$ on $a/b$ are shown. In the case of oblate semiellipsoids the graph $z_{min} - z_{max}$ versus $a/b$ is nonmonotonous (Fig.4c). It demonstrates the maximum at $a/b \approx 0.58$ and linear tendency to zero if $a/b \to 0$ (Fig.4c). In the case of oblate semiellipsoids the values of $|W_{min}|$ (Fig.4c) and $W_{max} - W_{min}$ (Fig.4d) decrease quickly as the ratio $b/a$ increases.

In the case of prolate semiellipsoid $z_{min} - z_{max}$ increases with an increase in the ratio a/b (Fig.4c) and, therefore, we can not assume converging of the extremes in the limiting case b/a→0. In this case, the elongated semiellipsoid can be approximately considered as a very long thin cylindrical tube. Considering the symmetry predicts the maximum of $W(z_q)$ at the tube center. If $z_q \to \pm\infty$, attraction between the uncharged tube and point charge is expected, hence, $W(z_q) \to 0$. As a result, the minimum of the function $W(z_q)$ has to exist, that requires $|W_{min}| \neq 0$ (see points in Fig.4b for a>b). According to Fig.4a, this minimum can be located near $z_q$=0. So, we can assume that the repulsion effect continues to exist in the case of elongation of an open semiellipsoid.

### 3c. Truncated spherical conductor

Figure 5a shows the uncharged truncated open spherical conductor of radius $R$ in the field of the point charge $q$, which can be located at or $z_q \epsilon (-R, +\infty)$. The coordinates of the sphere center and the hole are equal to zero and $z_h$, respectively. Figure 5b shows the curves $W(z_q)$ for various values of $z_h$. The repulsion effect takes place for $z_q \epsilon (z_{max}, z_{min})$. The coordinate $z_{max}$ and the value of the local maximum of $W(z_q)$ are equal to zero for all curves, because if $z_q$=0, charges on a thin-wall sphere are not induced. The coordinate $z_{min}$ and the absolute value $|W_{min}|$ of the local minumum strongly depend on the coordinate $z_h$ of the hole in the sphere (Figs.5b).

Figure 6a shows on a logarithmic scale $|W_{min}|$ as a function of the hole coordinate $z_h$. The value of $|W_{min}|$ increases with increasing $z_h$ in the interval $z_h \epsilon (-R, +R)$, growing most quickly when $z_h \approx -R$ and $z_h \approx R$. The inset in Fig.6a shows variations of the difference $z_{min} - z_h$ with $z_h$. If $z_h$ is close to $R$ or $-R$, then $z_{min}$ is very close to $z_h$ or 0.5R, respectively.

Figure 6b displays $|W_{min}|$ as a function of the hole radius $R_h$. In the case of small hole and $z_h \approx R$ (Fig.6b) or $z_h \approx -R$ (the inset in Fig.6b) the value of $|W_{min}|$ is maximal or minimal, respectively. The difference between these extreme values of $|W_{min}|$ is five orders of magnitude. The reason for such a big difference can be explained as follows. If $z_h \approx R$, an open hemisphere and a closed sphere are nearly identical. The energy of a point charge inside and outside a neutral conducting closed sphere is determined by Eqs.6 and 7, respectively, using the image method. These equations, illustrated by the curves in Fig.6c, predict that for a closed sphere, if $z_q \to R$, then $W(z_q) \to -\infty$. For the open truncated sphere the curve $W(z_q)$ with a local minimum $W_{min}$ at $z_{min} \approx z_h \approx R$ is shown in the inset in Fig.6c. In the limit transition $z_h \to R$, the continuous curve $W(z_q)$ for the truncated sphere turns into the diverged curve with two branches (Fig.6c). This can be the reason for the increase in $|W_{min}|$ by five orders of magnitude, if $R_h \to 0$ (Fig.6b).

$$W(z_q) = \frac{q^2}{2} \left( \frac{1}{R} - \frac{R}{\sqrt{z_q^4 + R^4 - 2z_q^2 R^2}} \right), \tag{6}$$

$$W(z_q) = \frac{q^2}{2} \left( \frac{R}{z_q^2} - \frac{R}{\sqrt{z_q^4 + R^4 - 2z_q^2 R^2}} \right). \tag{7}$$

Figure 6d illustrates the decrease, if $z_h<0$, and increase, if $z_h>0$ (see the inset in Fig.6d), of the repulsion force with decreasing $R_h$, by showing the ratio $(W_{max} - W_{min})/(z_{min} - z_{max})$ versus $R_h/(z_h + R)$ for the truncated open sphere. This ratio is reasonable estimation of the magnitude of the force in the interval $z_q \in (z_{max}, z_{min})$, where the repulsion occurs. In the case of a decrease in $R_h$, the ratio $(W_{max} - W_{min})/(z_{min} - z_{max})$ goes to zero, if $z_h<0$, and quickly increases, if $z_h>0$. In the case of $z_h>0$, $W_{max} - W_{min}$ increases with decreasing $R_h$, while $z_{min} - z_{max}$ decreases from 0.63R at $z_h$=0 down to zero at $z_h$=R. As a result, the repulsion force can grow significantly (the inset in Fig.6d). In the case of $z_h<0$, both values $W_{max} - W_{min}$ and $z_{min} - z_{max}$ decreases with decreasing $R_h$, but the former decreases more significantly (see Fig.6a). Note that $W_{max}$ and $z_{max}$ are always zero, $z_{min}$ varies from 0.63R at $z_h$=0 up to 0.5R at $z_h = -R$.

A similar dependency $(W_{max} - W_{min})/(z_{min} - z_{max})$ versus $b/a$ for the oblate semiellipsoid is shown in Fig.6d for comparison. The repulsive force gradually decreases with an increase in the $b/a$ ratio (Fig.6d), when $W_{max} - W_{min}$ vanish (Fig.4b) and the local minimum and maximum merge (Fig.4c). In the case of a truncated open sphere the local minimum also gradually disappears, if $z_h \to -R$, but the distance between the local extremes remains finite.

## 4. Equivalent model of repulsion effect for uncharged conductors

An equivalent physical model can be introduced that easily explains the repulsion effect for a neutral conductor in a point charge field. This model replaces the real charge distribution on the conductor with a system of a point charge $Q_+$ and a uniformly charged ring with a total charge $Q_-$ (Figs.7-9). Assuming that the positive point charge q is outside the neutral open hemisphere, the negative induced charges are located as close as possible to the charge q. The positive induced charges are located as far away from the charge $q$. Correspondingly, in our model $Q_-<0$ and $Q_+>0$. Due to the electroneutrality of the conductor the following relation is true: $Q_+ + Q_- = 0$. The electrostatic energy of the system can be found:

$$W(z) = \frac{qQ_-}{\sqrt{z^2 + R^2}} + \frac{qQ_+}{z + R}. \tag{8}$$

The comparison of (8) with the exact solution [Levin, 2011] is presented in Fig.7b. The absolute value of the induced charge $Q = Q_+ = |Q_-|$ depends on a distance between the point charge q and the hemisphere (the inset in Fig.7b). The geometric analysis of the charge distribution

in Fig.7a allows us to explain the physical cause of repulsion. The negative induced charges ($Q_-$) are closer to the point charge q than the positive induced charges ($Q_+$). However, the negative induced charges act on the point charge ineffectively because of a large angle θ to the axis of symmetry (Fig.7a). When the point charge q is near the origin, the angle θ is greatest. In this case the repulsive force ($F_{Q_+}$) from the positive induced charges $Q_+$ prevails over the attractive force ($F_{Q_-}$) from the negative induced charges $Q_-$ of the ring.

Our model allows to determine the coordinate of the local minimum $z_{min}$. The attractive force acting on a point charge $q$ along the symmetry axis $z$ due to the uniformly charged ring $Q_-$ is equal:

$$F_{Q_-} = -q\partial\frac{\varphi_{Q_-}}{\partial z} = zq\frac{Q_-}{((z-z_i)^2 + a_r^2)^{3/2}}, \tag{9}$$

where $\varphi_{Q_-}$ is the ring's potential (see Eq.(2)).

The repulsive force acting on the point charge $q$ along the $z$ axis due to the point charge $Q_+$ is equal:

$$F_{Q_+} = q\frac{Q_+}{(z-z_0)^2}. \tag{10}$$

In the case of the open semiellipsoid of revolution (Fig.8a), the total force on the point charge $q$ is given by

$$F_q = F_{Q_+} + F_{Q_-} = q\frac{Q}{(z+a)^2} - zq\frac{Q}{(z^2+b^2)^{3/2}}. \tag{11}$$

In the case of the truncated spherical conductor, the total force on the point charge $q$ (see Fig.9a):

$$F_q = F_{Q_+} + F_{Q_-} = q\frac{Q}{(z+R)^2} - (z-z_h)q\frac{Q}{(z^2 - 2zz_h + R^2)^{3/2}}. \tag{12}$$

The total force (11) and (12) acting on the point charge $q$ is zero, if $z=z_{min}$. As a result, we obtain nonlinear equations for finding $z_{min}$, in case of a semiellipsoid of revolution (Eq.13) and a truncated spherical conductor (Eq.14):

$$(a+z)^2 z = (b^2 + z^2)^{3/2} \tag{13}$$

$$(z-z_h)(z+R)^2 = (z^2 - 2zz_h + R^2)^{3/2} \tag{14}$$

The solutions of Eq.(13) and Eq.(14) are shown in Fig.8b and Fig.9b, respectively. In case of the semiellipsoid the values of zmin obtained by our numerical caluclations and by Eq(.13) are in good accordance if the ratio *a/b* is near 1 (Fig.8b). In case of the truncated sphere the difference between values of $z_{min}$, calculated numerically and by Eq.(14), are in good accordance for all values of $z_h$ (Fig.9b).

As shown in Fig.9b, numerically calculated values of $z_{min}$ tend to 0.5R for $z_h \to -1$. This specificity of the value $z_{min}=0.5$ can be shown from the condition $dz/dz_h = 0$ by using Eq.(14). Differentiation of the right and left sides of Eq.(14) and addition of similar terms, containing $dz$ and $dz_h$, result in Eq.(15):

$$\frac{dz}{dz_h} = \frac{\left((z+R)^2 - 3z\sqrt{z^2 - 2zz_h + R^2}\right)}{\left((z+R)^2 + (z-z_h)\left(2(z+R) - 3\sqrt{z^2 - 2zz_h + R^2}\right)\right)} = 0 \tag{15}$$

Setting the numerator in Eq.(15) to zero, we obtain Eq.(16):
$$(z + R)^2 - 3z\sqrt{z^2 - 2zz_h + R^2} = 0 \tag{16}$$

Inserting $z_h = -R$ in Eq.(16), we have Eq.(17):
$$(z + R)^2 - 3z(z + R) = 0 \tag{17}$$

Since $z \neq -R$, this equation is simplified:
$$z + R - 3z = 0$$

Hence, we obtain:
$$z_{min} = \frac{R}{2}$$

This result explains our numerical calculations of $z_{min}$ values, in the case when $z_h/R$ is close to $-1$ (Fig.9b).

## 5. Repulsion between charged open hemisphere and point charge

The last sections present the repulsion effect for uncharged conductors of various shapes in the electric field of a point charge. Is the repulsion effect possible for the same systems if the conductor is charged? Our methods allow to consider both uncharged and charged conductors, for example, the charged open hemisphere in the field of the point charge $q$. This system with the uncharged hemisphere has already been analyzed in section 3a. It is convenient to consider the electrostatic potential in the case of the charged hemisphere in the field of the point charge as the superposition of two functions. The first is the exact solution for the system with an uncharged hemisphere [Levin, 2011] (see Fig.2b). The second is our numerical solution for an isolated charged hemisphere in the absence of a point charge q. It is enough to calculate the second function only once, for $Q=1$. In the case of $Q\neq 1$, this function is simply multiplied by $Q$. Our numerical calculations of this function by the first method allow find the value of the electric capacity of the open hemisphere which is equal to 0.823. This value is very close to the exact result $C/R=(1/2+1/\pi)=0.818$ [Snow, 1954].

Figure 10a shows energy curves $W(z_q)$ in the case of $Q\neq 0$, where $Q$ is total charge of the hemisphere.

If $Q_n<Q<0$, where $Q_n = -0.052q$, there are three areas (I-III) corresponding to repulsive or attractive forces. The areas I, III and II correspond to attraction and repulsion, respectively. If $Q=Q_p$, then $W_{min}=W_{max}$ and $z_{min}=z_{max}$ (Fig.10b). The local maximum and minimum merge (the inset in Fig.10b) and the area II, corresponding to repulsion, disappears. If $Q_p>Q>0$, where $Q_p=0.023q$, there are four areas (I-IV), indicated for the upper curve in Fig.10a. The areas I, III and II, IV correspond to attraction and repulsion, respectively. If $Q=Q_p$, then $W_{min}=W*_{max}$ and $z_{min}=z*_{max}$ (Figs.10c,d). The local maximum and minimum merge and the area III, corresponding to repulsion, disappears.

Figure 11 shows the general view of transformation of the extremes and their coordinates $z_{max}$, $z_{min}$ and $z*_{max}$ depending on of the hemisphere's charge $Q$. The minimum and the right maximum (see Fig.10a) exist in the finite interval of $Q$: $(Q_n, Q_p)$ and $(0, Q_p)$, respectively. The left maximum (see Fig.10a) occurs for all values of $Q$ exceeding $Q_n$.

Thus, the repulsion effect, considered in our paper, for the charged open hemisphere and the positive point charge exist only in a limited range of the total charge of the hemisphere: $Q_n<Q<Q_p$. In the case of $Q<Q_n<0$, only attraction between bodies occurs at all distances. In the case of $Q_p>Q>0$, repulsion occurs at two different intervals of $z_q$, at large and short distances between the sphere and the point charge.

## 6. Conclusions

Calculations and analysis of the electrostatic field of several types of axially symmetric uncharged thin-walled conductors with an open cavity in the field of a point charge are carried out. It is shown that the non-obvious repulsion effect of a charged and uncharged conductor, known from [Levin, 2011], significantly depends on the geometric shape of an uncharged conductor. For a conductor in the form of an oblate semi-ellipsoid, an increase in eccentricity leads to a smooth disappearance of the repulsion effect by narrowing the spatial area of its existence, removing it from the conductor and reducing the repulsion force. For a conductor in the form of a truncated sphere, the position of the hole relative to the center of the sphere determines both the possibility of the effect itself, as well as the position of the repulsion area and the magnitude of the repulsion force. It is shown that for an almost closed sphere with a small hole diameter, the magnitude of the repulsive force of a point charge from the sphere can be significant, an equivalent model is proposed in which continuously distributed induced charges are replaced by a discrete system of charges: a uniformly charged narrow ring and a point charge opposite in sign to the charge of the ring. The model allows us to calculate for the electrostatic energy the position of the local minimum, the existence of which is necessary for the manifestation of the repulsion effect. For a conductor in the form of a hemisphere, the effect of the appearance of a charge on an initially neutral conductor on the modification and disappearance of the repulsion effect is investigated. The charge range of the hemisphere is found for which the repulsion effect does not disappear. For a weakly charged hemisphere with a charge sign the same as that of a point charge, alternating two regions of attraction and two regions of repulsion were found, depending on the distance from the point charge to the hemisphere. 

**Declaration of competing interest**

The authors declare that they have no known competing financial interests or personal relationships that could have appeared to influence the work reported in this paper.

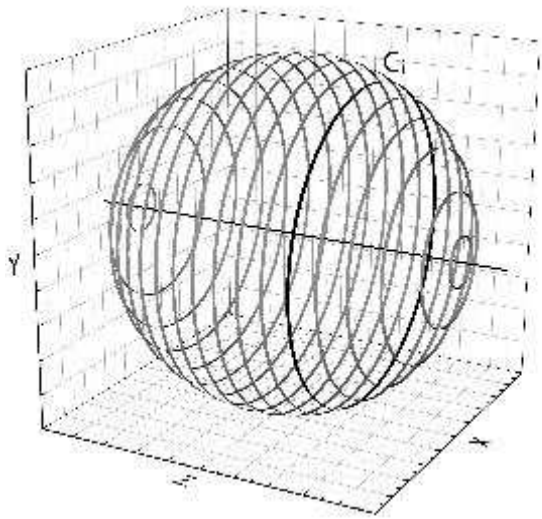 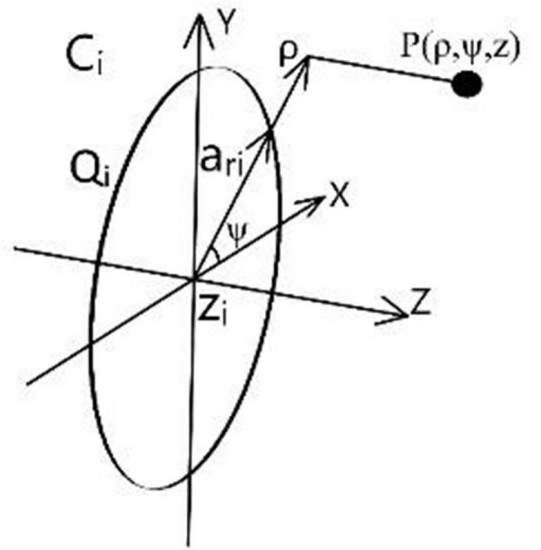

(a) (b)

Fig.1. (a) Representation of a sphere, as an example of an axial symmetric conductor, in the form of a set of thin coaxial rings $C_i$, $i=1, 2,..., N$. (b) Scheme for calculations of $E$ and $\varphi$ produced by $C_i$ at $P(\rho,\psi,z)$ according to Eqs.(1), (2); $a_{ri}$, $z_i$ and $Q_i$ denote the radius, center's coordinate and total charge of $C_i$, respectively.

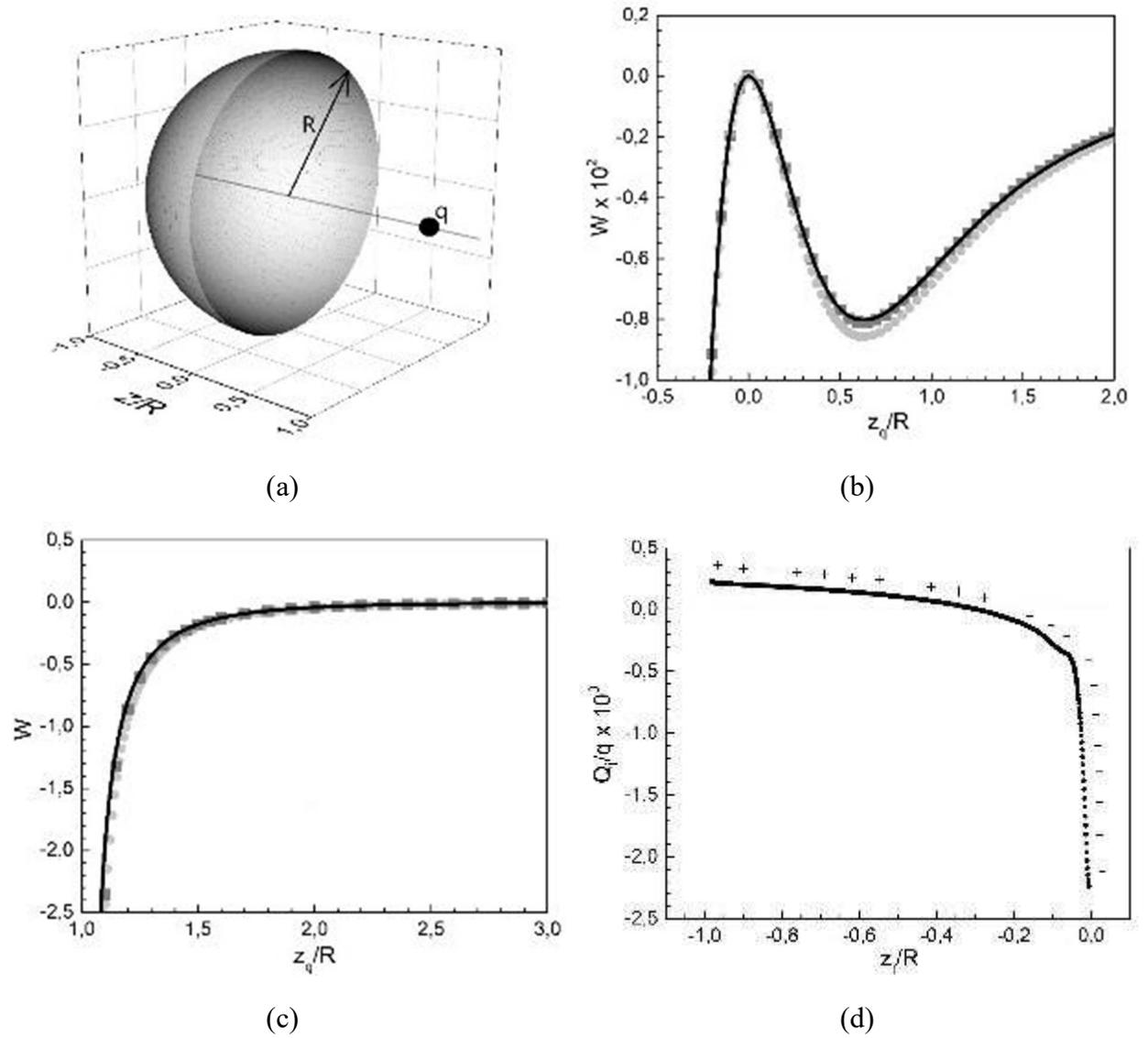

Fig.2. Sketch of a system containing a neutral conducting open hemisphere and a point charge $q$ (a). The energy $W$ as a function of $z_q$ for an interaction between the point charge and the neutral open hemisphere (b) or the neutral closed sphere (c). The energy $W(z_q)$ is normalized to $q^2/R$, where $R$ is the radius of the sphere and hemisphere; $z_q$ and $q$ are a coordinate and a charge of the point charge, respectively. Solid lines represent the exact solutions; symbols refer to results of the numerical calculations. Results of the first and second methods are shown by dark and light gray squares, respectively. (d): The electric charges $Q_i$ versus $z_i$ for $C_i$, $i=1,2,3...N$ ($N=2000$) (see Fig.1) corresponding to the local minimum of $W$ at $z_q = 0.63$. The symbols "+" and "−" refer to positive and negative induced charges, the total values of which are equal to $\pm 0.09q$, respectively.

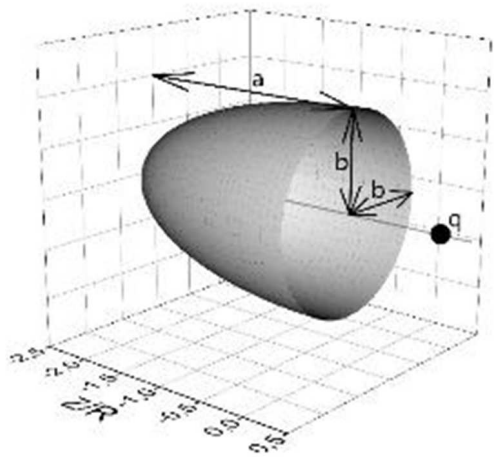
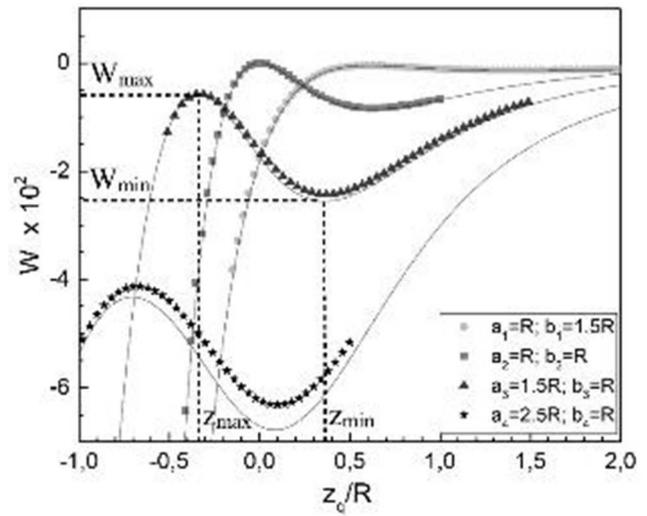

(a)                  (b)

Fig.3. (a) Sketch of a system containing a point charge $q$ and a neutral conducting open semiellipsoid of revolution. (b) The interaction energy $W$ of the system versus the coordinate $z_q$ of the point charge. The value of $W$ is normalized to $q^2/R$, where $R=a$ if $a<b$ and $R=b$ if $b<a$. If $a=b=R$, then the semiellipsoid is identical to the hemisphere. Symbols and solid lines denote the results of the first and second calculation methods, respectively. The dashed lines refer to local extremes $W_{min}$, $W_{max}$ at the $z_{min}$, $z_{max}$, respectively. The accuracy of determining $z_{max}$ and $z_{min}$ depends on the step size, which was usually 0.01R.

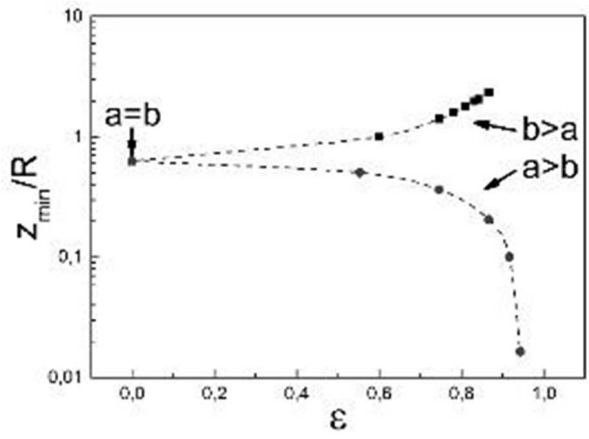
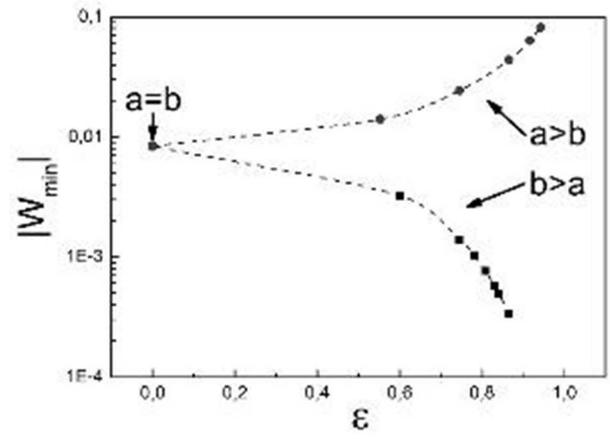

(a)                                            (b)

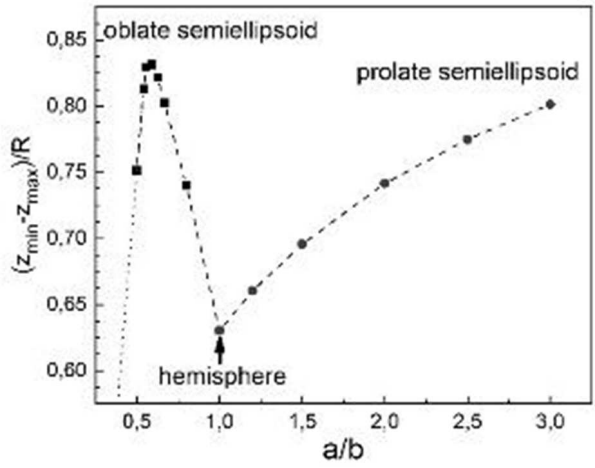
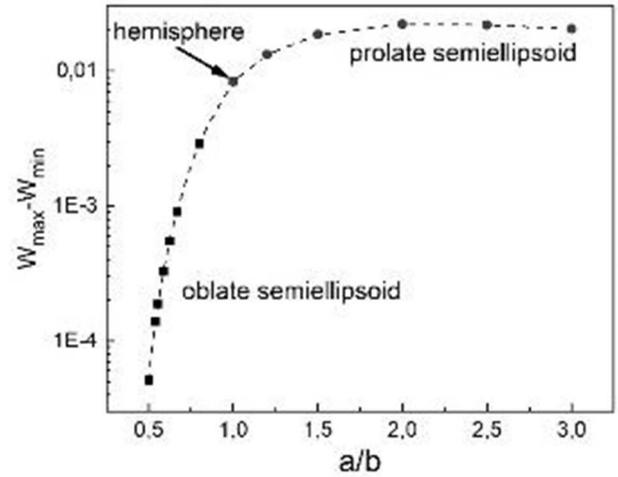

(c)                                            (d)

Fig.4 Variations of $z_{min}$ (a) and $|W_{min}|$ (b), $z_{min} - z_{max}$ (c) and $W_{max} - W_{min}$ (d) as functions of the eccentricity $\varepsilon$ and the ratio a/b, respectively, for perfectly conducting open semiellipsoid. The values $W_{min}$ and $W_{max}$ are normalized to $q^2/R$ (see the legend of Fig.3). The cases of oblate and prolate semiellipsoids are indicated. The results were obtained by the first method. The dashed lines are the guides for eye. The dot line is the straight line that intersects the coordinate origin.

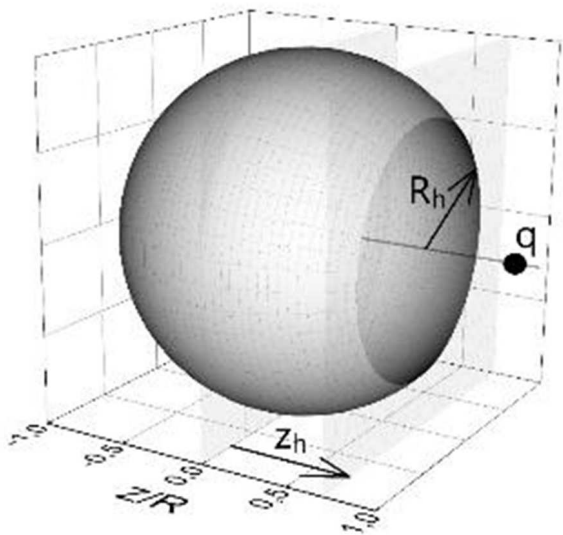 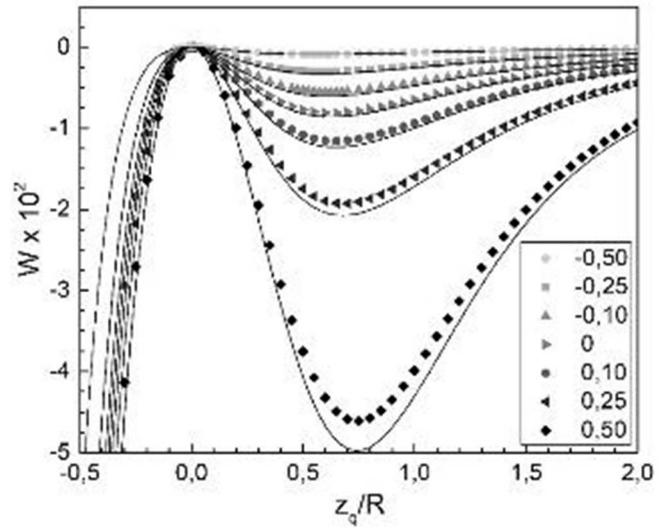

(a)                                            (b)

Fig.5. (a) Sketch of a system containing the neutral conducting truncated open sphere and the point charge $q$. The radius of the hole is denoted by $R_h$. (b) The interaction energy $W$ versus $z_q$, where $z_q$ is the point charge coordinate. The values of $W$ are normalized to $q^2/R$, where $R$ is the sphere radius. Numbers in the insert indicate the ratio $z_h/R$ for different curves, where $z_h$ is the hole coordinate. The energy curve for an open hemisphere is corresponded to $z_h=0$. Symbols and lines denote results of the first and second calculation methods, respectively.

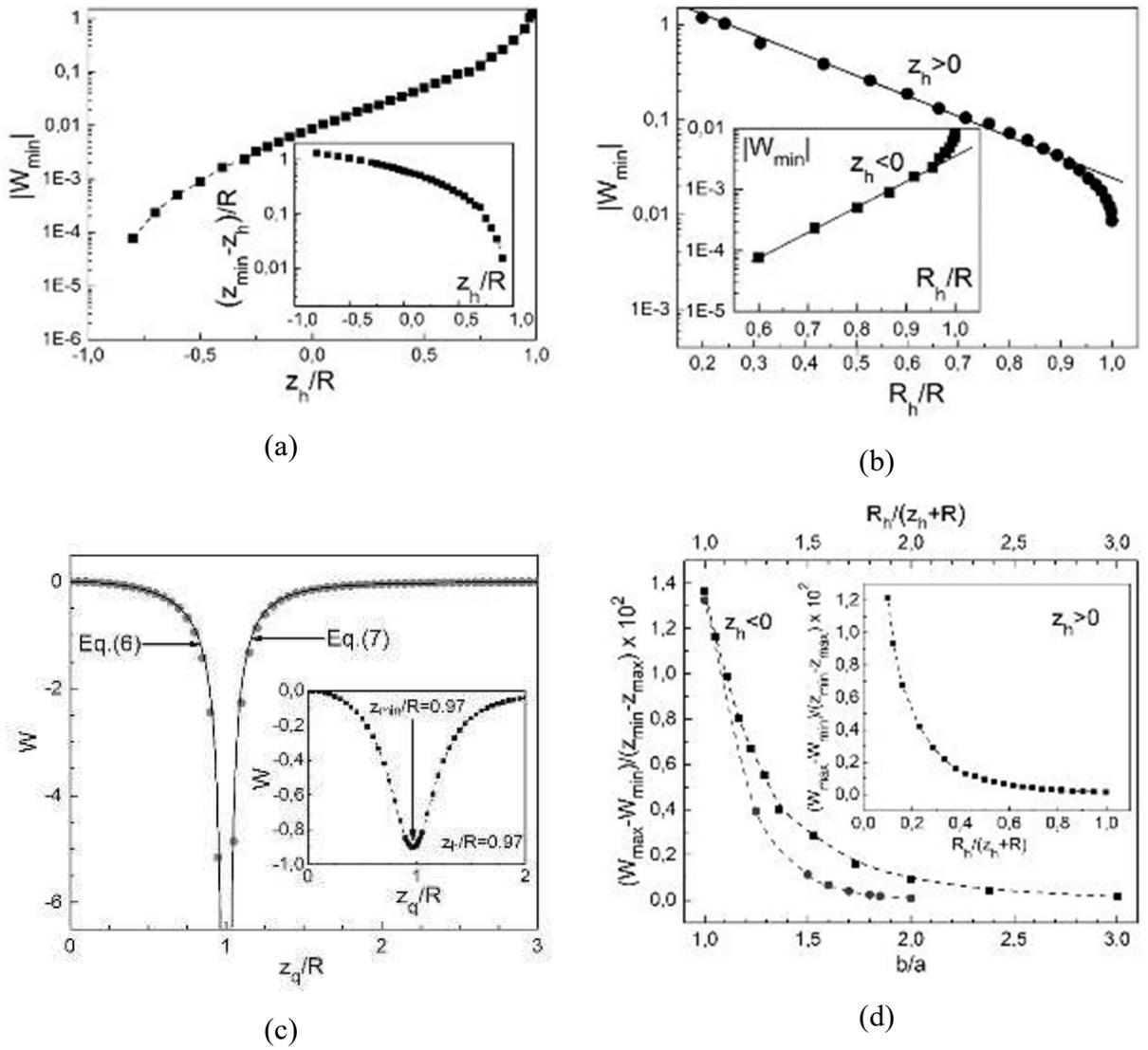

Fig.6. (a) and (b) Variations of $|W_{min}|$ for the truncated open sphere (Fig.5a) as functions of $z_h$ and $R_h$ (for $z_h>0$), respectively. The insets in (a) and (b) show $z_{min} - z_h$ versus $z_h$ and $|W_{min}|$ versus $R_h$ (for $z_h<0$), respectively. Note the logarithmic scale for the vertical axes. Solid lines are guides for eye. (c) The energy $W$ as a function of $z_q$ for the sphere (Fig.1a). Symbols are results of our first numerical method, solid lines are drawn using Eq.6 (for $z_q<R$) and Eq.7 (for $z_q>R$). Inset shows $W(z_q)$ for $z_h=0.97$. (d) The ratio $(W_{max} - W_{min})/(z_{min} - z_{max})$ as a function of $R_h/(z_h+R)$ for the truncated open sphere and as a function of $b/a$ for the oblate semiellipsoid (Fig.3a), respectively.

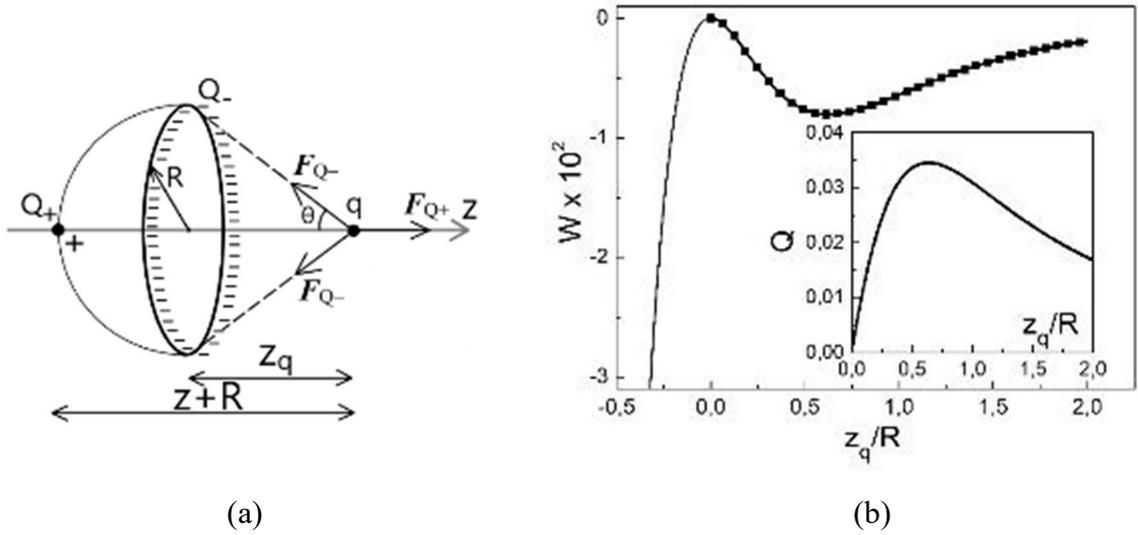

(a)                                         (b)

Fig.7. (a) Representation of the neutral hollow hemisphere in the field of the point charge q as the uniformly charged ring $Q_-$ and the point charge $Q_+$. The symbols "+" and "−" refer to positive and negative induced charges, respectively. $R$ is the radius of the hemisphere; $z_q$ is the coordinate of the point charge. Arrows $F_{Q_-}$ and $F_{Q_+}$ denote directions of the total forces on the point charge $q$ due to the interaction with the induced charges $Q_-$ and $Q_+$, respectively; θ is the angle between $F_{Q_-}$ and z axis; (b) The energy $W$, normalized to $q^2/R$, as a function of $z_q$. Solid line and squares represent the exact solution [Levin, 2011] and Eq.8, respectively. Inset shows $Q = Q_+ = |Q_-|$ as a function of $z_q$.

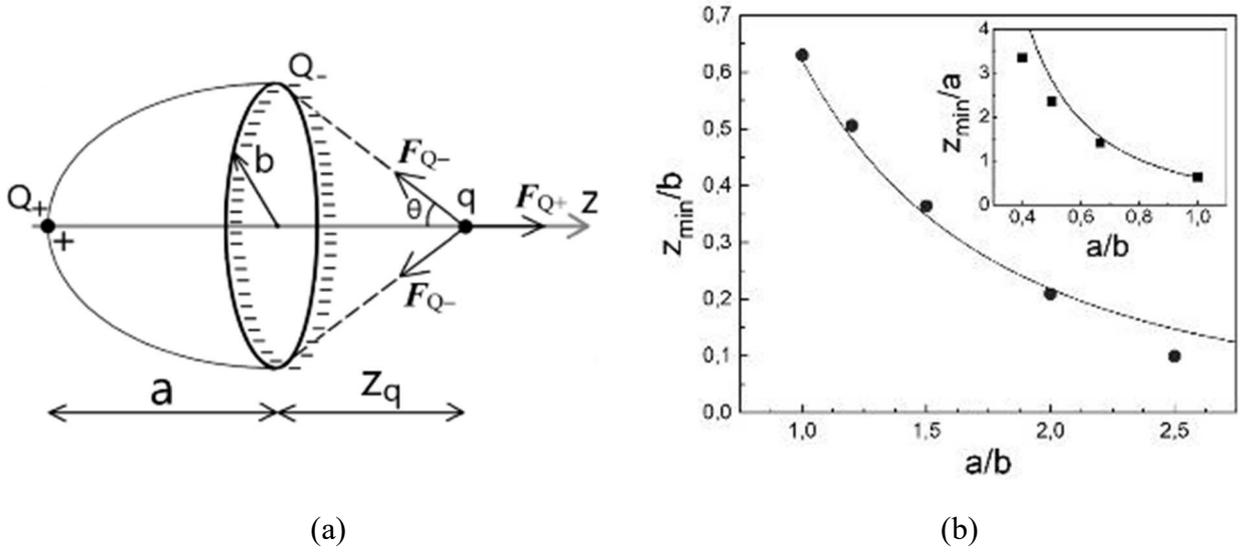

(a) (b)

Fig.8. (a) Representation of the neutral hollow semiellipsoid of revolution in the field of the point charge q as the uniformly charged ring $Q_-$ and the point charge $Q_+$ (see notations in Fig.7). Letters $a$, $b$ denote the semi-axes of the semiellipsoid. (b) The dependencies of $z_{min}$ for a semiellipsoid: prolate (on a/b, circles) and oblate (on b/a, squares in the inset). Solid lines represent the numerical solutions of Eq.(13).

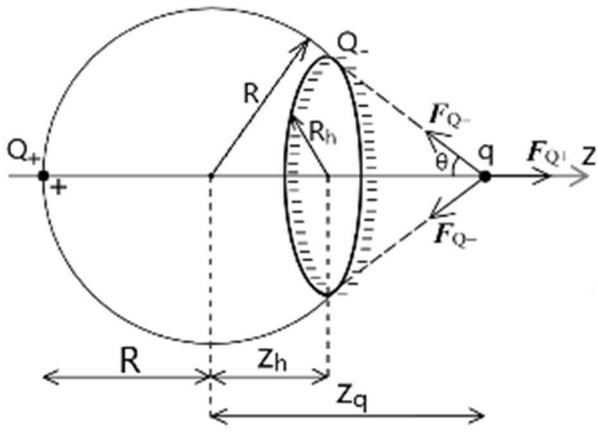 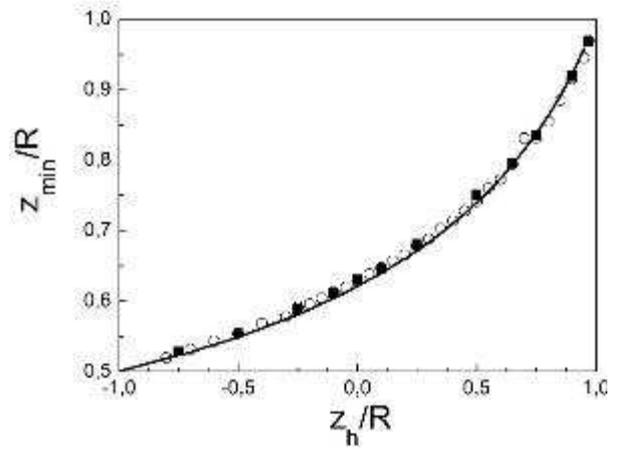

(a)                          (b)

Fig. 9. (a) Representation of the neutral truncated spherical hollow conductor in the field of the point charge q as the uniformly charged ring $Q_-$ and the point charge $Q_+$ (see notations in Fig.7) (b) The dependence of $z_{min}$ on $z_h$. Solid line represents the numerical solutions of Eq.(14). The results of the first and second methods are shown by squares and circles, respectively.

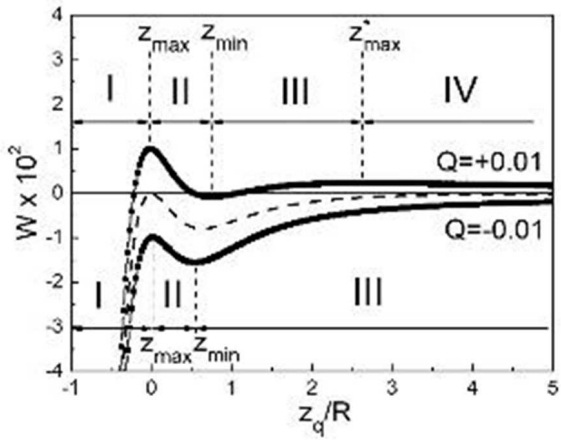
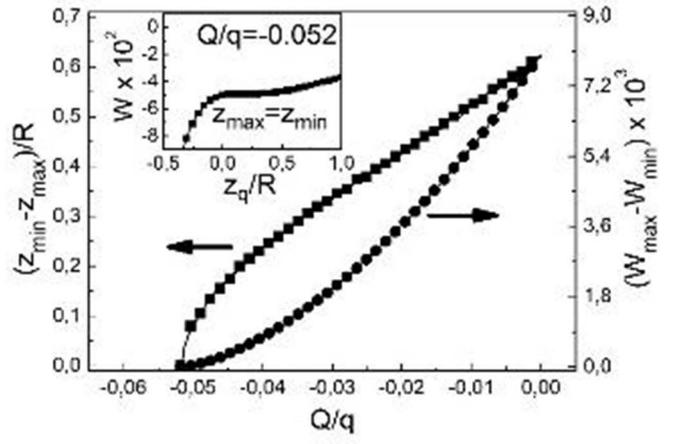

(a)                              (b)

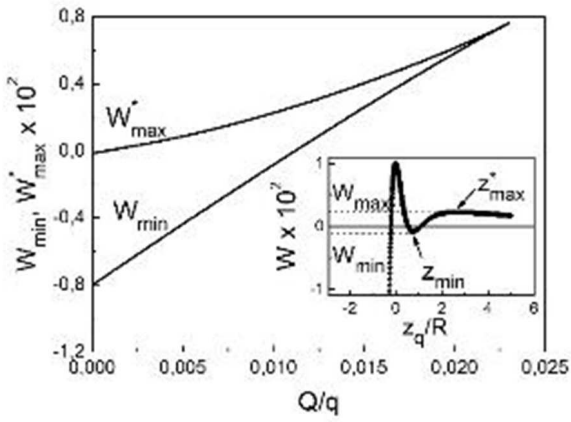
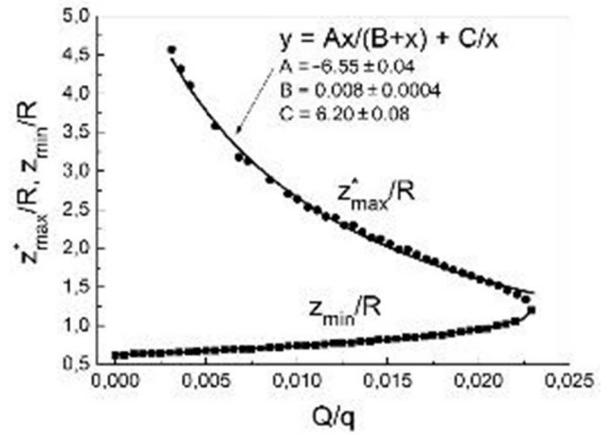

(c)                              (d)

Fig.10. (a) The energy W of the interaction between the charged hemisphere and the point charge q as a function of $z_q$. $Q$ is the total charge of the hemisphere. The value of $W$ is normalized to $q^2/R$, where $R$ is the radius of the hemisphere. The upper and lower curves correspond to $Q/q=0.01$ and $Q/q=-0.01$. The dashed line represents the exact solution [Levin, 2011] for $Q=0$. The numbers I–IV indicate areas of different signs of $F_q$ (Eq.12) in accordance with the slope of the curve $W(z_q)$. The areas I and III correspond to attraction, the areas II and IV correspond to repulsion. Vertical dashed lines indicate the local extremes $W_{min}$, $W_{max}$, $W^*_{max}$ at the $z_{min}$, $z_{max}$, $z^*_{max}$, respectively. (b) Variations of $z_{min}$-$z_{max}$ (left curve) and $W_{max}$-$W_{min}$ (right curve) versus $Q<0$. Insert shows the energy curve for $Q=-0.052$ when $z_{min}=z_{max}$ and the local minimum and maximum (see lower curve in (a)) coincide and disappear. (c) Variations of $W_{min}$ and $W^*_{max}$ versus $Q>0$. Insert shows the curve $W(z_q)$ for $Q=0.01$ on a large scale. (d) Variations of $z_{min}$ and $z^*_{max}$ versus $Q>0$. Solid curve was calculated using NLSF procedure in Origin 7.0.

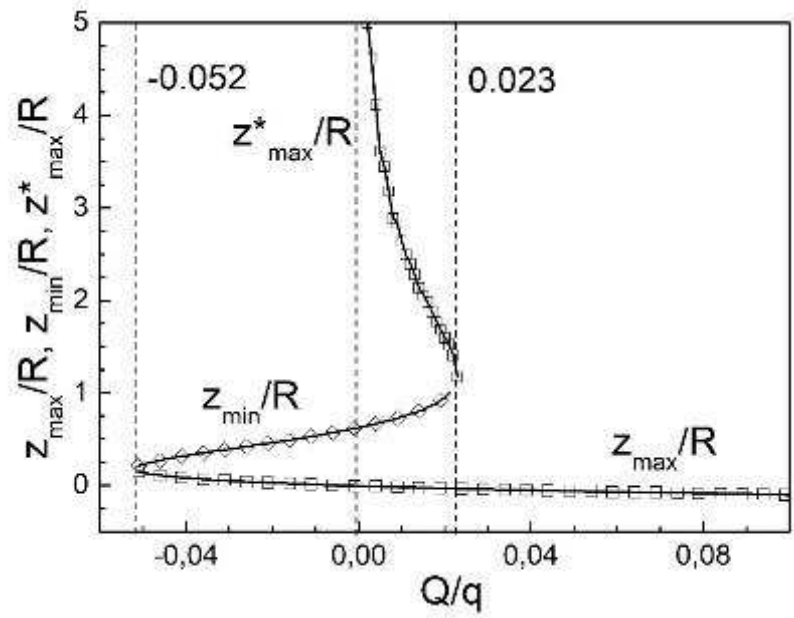

Fig.11. Variations of $z_{max}$, $z_{min}$ and $z^*_{max}$ (see Fig.10a) versus $Q$. The $Q$ values of arisement and emergence of the extremes, $Q_n = -0.052$, zero and $Q_p = 0.023$, are indicated by the dashed lines.